\documentclass[twocolumn,english,showpacs]{revtex4}
\usepackage[latin9]{inputenc}
\usepackage{amssymb}
\usepackage{graphicx}
\usepackage{amsmath}
\usepackage{color}
\usepackage{float}
\usepackage{mathrsfs}
\usepackage{indentfirst}
\usepackage{textcomp}
\usepackage{comment}
\usepackage{mathtools}

\newcommand{\COMMENTED}[1]{}

\begin{document}

\title{Ultracold atoms in a square lattice with spin-orbit coupling: \\
Charge order, superfluidity, and topological signatures}

\author{Peter Rosenberg}
\author{Hao Shi}
\author{Shiwei Zhang}

\affiliation{Department of Physics,
             The College of William and Mary,
             Williamsburg, Virginia 23187}

\begin{abstract}
We present  an \emph{ab initio}, numerically exact study of attractive fermions in square lattices with Rashba spin-orbit coupling. 
The ground state of this system is a supersolid, with co-existing charge and superfluid order. 
The superfluid is composed of both singlet and triplet pairs induced by spin-orbit coupling. We perform
large-scale calculations using auxiliary-field quantum Monte Carlo to provide the first full, quantitative description of the charge, 
spin, and pairing properties of the system. In addition to characterizing the exotic physics, 
our results will serve as essential high-accuracy benchmarks for the intense
theoretical and especially experimental efforts in ultracold atoms to realize and understand an expanding variety of 
quantum Hall and topological superconductor systems.
\end{abstract}


\maketitle

Exotic states of matter, including high-$T_c$ superconductivity and topological phases, have long been a focus of condensed 
matter physics. However, many of these behaviors are challenging to detect and 
characterize in real materials, where fixed structural and electromagnetic properties typically provide limited access to the parameter space. Modern cooling and trapping techniques in ultracold atoms, combined with optical lattice potentials \cite{reviewBloch,reviewLewenstein} and artificial gauge fields \cite{artificialgaugefieldJaksch,PhysRevLett.95.010403,PhysRevLett.108.225303,PhysRevLett.107.255301,PhysRevLett.111.185302,GerbierArtifcialGaugeFields} have opened a new window onto these novel phases, which can be explored across broad parameter regimes well beyond those available to experiments with real materials. 

Ultracold atoms in optical lattices thus provide
access to clean, tunable systems in which to study the combination of strong interactions and spin-orbit coupling (SOC) with extreme precision. These systems display a number of exotic phases. A thorough understanding, and high-accuracy characterization, of these phases will have important implications across the fields of spintronics, and quantum computation and information \cite{AliceaMajoranaReview, Review_anyons_quantum_comp, PhysRevLett.104.040502}, among others.

The importance of high-accuracy numerical benchmarks has grown considerably with the recent advent of Fermi gas microscopes, capable of performing site-resolved measurements of lattice fermions \cite{PhysRevLett.116.235301, Greif953, Cheuk2015, Haller2015, Parsons2015}.   
This expansive experimental horizon, and the promise of high-accuracy measurements of these disorder-free and 
finely tunable systems, has attracted considerable theoretical interest \cite{PhysRevLett.109.085303,determinant_QMC_oplatt,CDMFT_2DFG_SOC,Rasbha2DLatticeMF,BHM_SOC}. Complementary experimental and theoretical
efforts will help shed light on the novel physics realized in these systems, which combine strong interaction, exotic pairing and superconductivity, as well as topological effects. 

Despite this rapidly growing interest, the ground state properties of lattice fermions with attractive interactions and SOC, 
a candidate system for the realization of many novel phases, remain largely uncharacterized beyond mean-field theory
 \cite{Rasbha2DLatticeMF}. Many-fermion systems are notoriously challenging to treat theoretically or computationally. 
Applying several advances in auxiliary-field quantum Monte Carlo (AFQMC) simulations \cite{Lecture-notes, BSS, Koonin}, 
including accelerated sampling \cite{2DFG_PRA} with force bias \cite{PRL-phaseless}, control of Monte Carlo variance 
\cite{inf_var}, and treatment of SOC in AFQMC \cite{2DFG_SOC_AFQMC}, we show that systematic and high-precision 
numerical data can be obtained on the ground-state properties of this remarkable system. The spin-balanced situation, when no Zeeman 
field is present, preserves time-reversal symmetry such that the calculations can be made free of the fermion sign problem
\cite{2DFG_SOC_AFQMC,Majorana-sign}, and the results here are unbiased and numerically exact.
This work thus serves as an illustration of the power of state-of-the-art computational approaches for 
exotic, strongly correlated quantum systems, in a system which is on the verge of experimental realization.

Our results allow a quantitative 
description of attractive fermions with Rashba SOC in a
two-dimensional (2D) optical lattice, which displays a supersolid phase \cite{supersolidChester, supersolidAndreevLifshitz,
supersolidLeggett, 2DHubbardNegU_Scalettar,2DHubbardNegU_Moreo} with both singlet and triplet pairing, 
and topological signatures. We elucidate the unique pairing properties of the system, and their connections to the spin 
and momentum distributions. The interplay between pairing and the charge order at half-filling is characterized.
Moreover, we examine the edge currents to explore the emergence of topological behavior, 
in the context of Majorana edge modes which are of strong current interest.

 The Hamiltonian for ultracold fermions in a 2D optical lattice with Rashba SOC is written,
\begin{align}
\hat{H}=&\sum_{\mathbf{k}\sigma}\varepsilon_{\mathbf{k}}c^\dagger_{\mathbf{k}\sigma} c_{\mathbf{k}\sigma} 
+ \sum_{\mathbf{k}} \left(\mathcal{L}_\mathbf{k} c^\dagger_{\mathbf{k}\downarrow} c_{\mathbf{k}\uparrow} 
+ \textrm{h.c.}\right) \notag
\\& +\sum_\mathbf{i} U n_{\mathbf{i}\uparrow}n_{\mathbf{i}\downarrow},
\end{align}
with $\varepsilon_\mathbf{k} = -2t(\cos k_x+\cos k_y)$, and $\mathcal{L}_\mathbf{k} = 2\lambda(\sin k_y-i\sin k_x).$
The parameter $t$, set to unity throughout this work, determines the strength of nearest-neighbor hopping, the
parameter $\lambda$ controls the strength of SOC,
and the parameter $U$ ($<0$) determines the strength of the on-site attractive 
interaction, 
with $n_{i\sigma}=c^\dagger_{i\sigma} c_{i\sigma}$ denoting the density operator in real-space 
on site $i$ with spin-$\sigma$ ($=\uparrow$ or $\downarrow$).
We will focus mostly on half-filling in this work, with a total number of atoms equal to the number of lattice sites.

\begin{figure}[!h]
\includegraphics[width=\columnwidth]{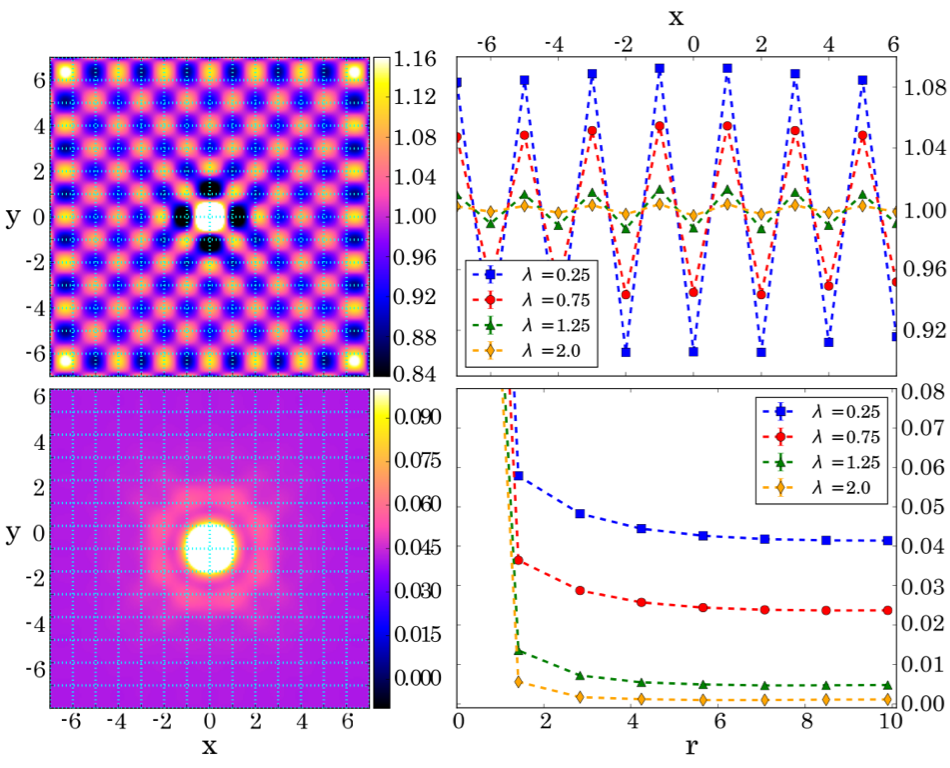}
\caption{Density-density (upper panel) and pair-pair (lower panel) correlation functions. The left column is for 
one SOC strength $\lambda=0.25$, while the
right column shows the evolution versus $\lambda$.
The density plots on the left show the correlations for $i$ running through the entire periodic supercell ($14\times 14$).
The upper right panel shows $D(i,0)$ for a slice of sites $i$ along $i_y=-3$,
while the lower right panel shows $P_{1s}(i,0)$ for a slice along the diagonal, $i_x=i_y$.}
\label{fig:ninj_didj_vs_lbda}
\end{figure} 

Although the existence and stability of a supersolid (superconducting charge-density-wave) state in the attractive Hubbard
model has been well understood \cite{2DHubbardNegU_Scalettar,2DHubbardNegU_Moreo},
the properties of this phase in the presence of Rashba SOC remain largely uncharacterized. 
Here we provide a precise characterization of this phase, and study its properties and their interplay with SOC strength. 
We examine the charge order by measuring the density-density correlation
function $D(i,j)= \langle n_i n_j\rangle $, with $n_i=(n_{i\uparrow}  + n_{i\downarrow})$.
(Expectation values denoted by angle brackets are always taken with respect to the many-body ground state
in this work.) Several quantities are measured to probe superfluidity, including the $s$-wave pair-pair correlation
$P_{1s}(i,j)= \langle \Delta^\dagger_{i} \Delta_{j} \rangle$ where 
$\Delta^\dagger_{i}=c^\dagger_{i\uparrow} c^\dagger_{i\downarrow}$ creates a  singlet-pair 
 at lattice site $i$. In periodic supercells we average over the reference site $j$ and will label it 
 as '0' for convenience.

A characteristic example of the supersolid state is shown in Fig.~\ref{fig:ninj_didj_vs_lbda}. Plotted in the upper left
corner is the density-density correlation function, which provides an illustration of the charge density wave (CDW).
A clear checkerboard pattern is seen, with a persistent alternating order. Sites that belong to the same sub-lattice as the 
reference have above average occupation, while those on the other sub-lattice have  below average occupation.
The amplitude of the CDW is essentially constant across the lattice. We have verified with 
calculations on larger supercell sizes that the long-range values of the CDW is independent of size,
providing strong evidence of long-range order. Plotted below the density-density correlation is the pair-pair correlation function. 
In this case, the presence of long-range superfluid order is evident from the significant magnitude of the pair-pair correlation across 
the lattice.

The effects of increasing SOC strength on these correlations are illustrated in the right column
of Fig.~\ref{fig:ninj_didj_vs_lbda}, which shows
slices of the density-density correlation in the top panel, and the pair-pair correlation in the bottom panel. 
The correlations evolve in equal proportion to each other with increasing SOC, which is a reflection of 
the underlying particle-hole symmetry. For small values of $\lambda$ the supersolid state is robust, with charge and 
superfluid orders of sizable magnitude. As SOC increases, these orders are diminished. However, even at strong SOC, 
such as $\lambda=2.0$, both remain finite.
 
\begin{figure*}[!ht]
\includegraphics[width=0.9\textwidth]{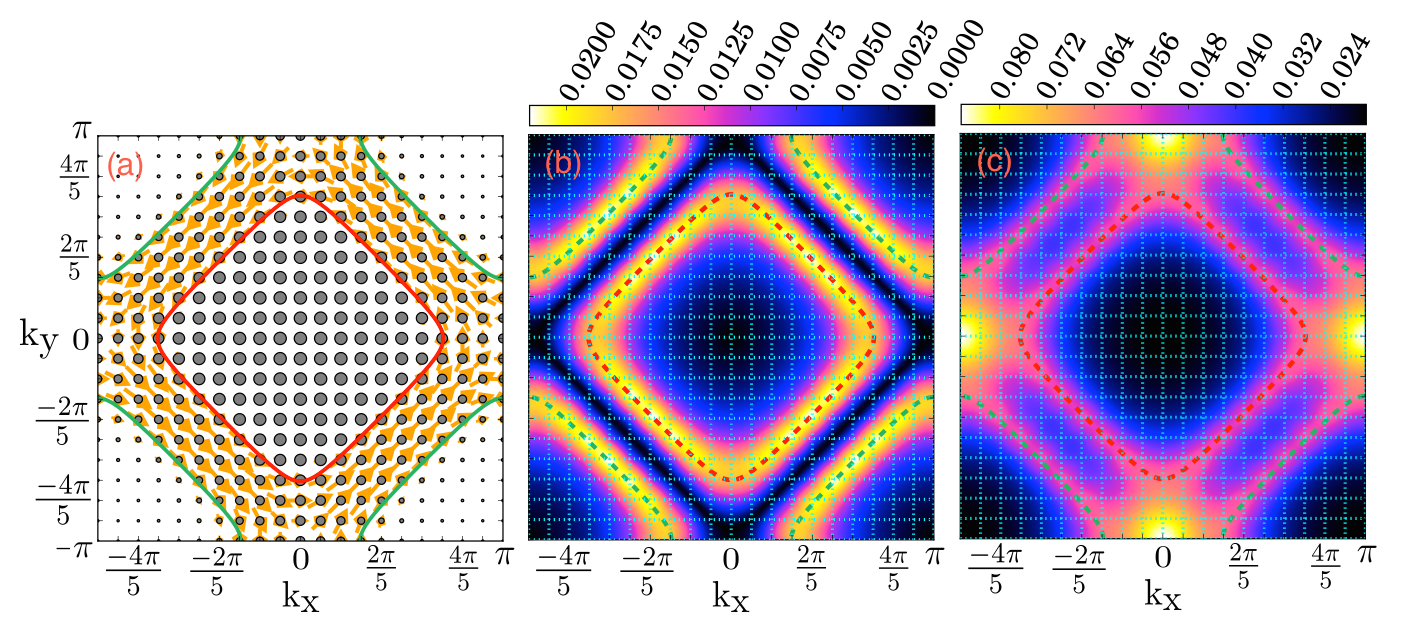}
\caption{Spin distribution and momentum-space pair wavefunctions. In (a) the arrow length and direction give the magnitude and orientation of the spin expectation at each lattice momentum, and the dot size is proportional to total occupation. The amplitude of the triplet and singlet parts of the pairing wavefunction are plotted in (b) and (c), respectively, with the non-interacting Fermi surfaces indicated by the dashed curves.
The system is a $20\times 20$ periodic supercell.
}
\label{fig:psi_k_spin_l0.5}
\end{figure*}

The nature of pairing is altered in a fundamental way by the presence of SOC.
The singlet and triplet channels are mixed, resulting in a pair wavefunction with both singlet and 
triplet components, $\Psi_p = \psi^s_p + \psi^t_p$.
We investigate pairing via construction and diagonalization of the two-body density matrix
in momentum space,
\begin{equation}
\COMMENTED{
\rho_2(\mathbf{k},\sigma;\mathbf{q},\sigma^\prime)=\langle
\Delta^\dagger_\sigma\left(\mathbf{k}\right)
\Delta_{\sigma^\prime}\left(\mathbf{q}\right) \rangle,
}
\rho_2(\mathbf{k},\chi;\mathbf{q},\chi^\prime)=\langle
\Delta^\dagger_\chi\left(\mathbf{k}\right)
\Delta_{\chi^\prime}\left(\mathbf{q}\right) \rangle,
\end{equation}
with $\chi=s$ or $t_\uparrow$ or $t_\downarrow$. The singlet and triplet pairing operators are
\begin{align}
\Delta^\dagger_s&=\frac{1}{\sqrt{2}}\left(c^\dagger_{\mathbf{k}\uparrow}c^\dagger_{\mathbf{-k}\downarrow}
-c^\dagger_{\mathbf{k}\downarrow}c^\dagger_{\mathbf{-k}\uparrow}\right),\\\notag
\Delta^\dagger_{t_\uparrow}&=c^\dagger_{\mathbf{k}\uparrow}c^\dagger_{\mathbf{-k}\uparrow},\\\notag
\Delta^\dagger_{t_\downarrow}&=c^\dagger_{\mathbf{k}\downarrow}c^\dagger_{\mathbf{-k}\downarrow}\,,
\end{align}
while the third component of the triplet pairing vanishes by symmetry. 
The leading eigenvalue, $N_c$,  of the $3N\times 3N$ matrix $\rho_2$ gives the condensate fraction of the pairs \cite{Yang1962}:
 $n_c\equiv N_c/N$. The corresponding eigenstate gives the pair wavefunction.
 
The structures of the pair wavefunctions,  $\psi^s_p({\mathbf k})$ and $ \psi^t_p({\mathbf k})$,
are illustrated in Fig.~\ref{fig:psi_k_spin_l0.5}. To better understand their physical origin,
we also compute the momentum distributions, which are shown in the left panel. 
At each ${\mathbf k}$, the total occupancy is given by $\langle n_{\mathbf{k}}\rangle=
\langle c^\dagger_{\mathbf{k}\uparrow} c_{\mathbf{k}\uparrow} + 
c^\dagger_{\mathbf{k}\downarrow} c_{\mathbf{k}\downarrow}\rangle$.
The expectations of the spin components are
 $\langle S^x_{\mathbf{k}} \rangle= \langle c^\dagger_{\mathbf{k}\uparrow} c_{\mathbf{k}\downarrow}
 +c^\dagger_{\mathbf{k}\downarrow} c_{\mathbf{k}\uparrow}\rangle/2$, 
and $\langle S^y_{\mathbf{k}} \rangle= \langle c^\dagger_{\mathbf{k}\uparrow} c_{\mathbf{k}\downarrow}
 -c^\dagger_{\mathbf{k}\downarrow} c_{\mathbf{k}\uparrow}\rangle/2i$,
with $\langle S^z_{\mathbf{k}} \rangle=0$ by symmetry here. 
Thus the magnitude and orientation of $\langle{\mathbf S}_{\mathbf{k}} \rangle$ can be determined.
In addition, the non-interacting Fermi surfaces are shown in the figure for reference, which 
are easily obtained from the 
two helicity branches of the
dispersion relations, $\varepsilon^\pm_\mathbf{k}=-2t\left(\cos k_x + \cos k_y\right)\pm 2\lambda
\sqrt{\sin^2 k_x + \sin^2 k_y}$.

The pair wavefunctions illustrate that triplet pairing occurs across the entirety of both Fermi surfaces (one for each helicity band). Singlet pairing also occurs across both Fermi surfaces, but with SOC it becomes strongly peaked near the four band-touching points, which occur at the half-filling Fermi level at $\mathbf{k} = (\pm\pi,0), (0,\pm\pi)$. As indicated by the spin and momentum distributions, these points have no excess spin, but are not fully occupied. They are also the only points in momentum-space that are surrounded by almost anti-parallel spins. 
Excitation of these neighboring anti-parallel spins into the momentum states at the band-touching points dramatically promotes singlet pairing across the Fermi surface at these points.

\begin{figure}[!ht]
\includegraphics[width=\columnwidth]{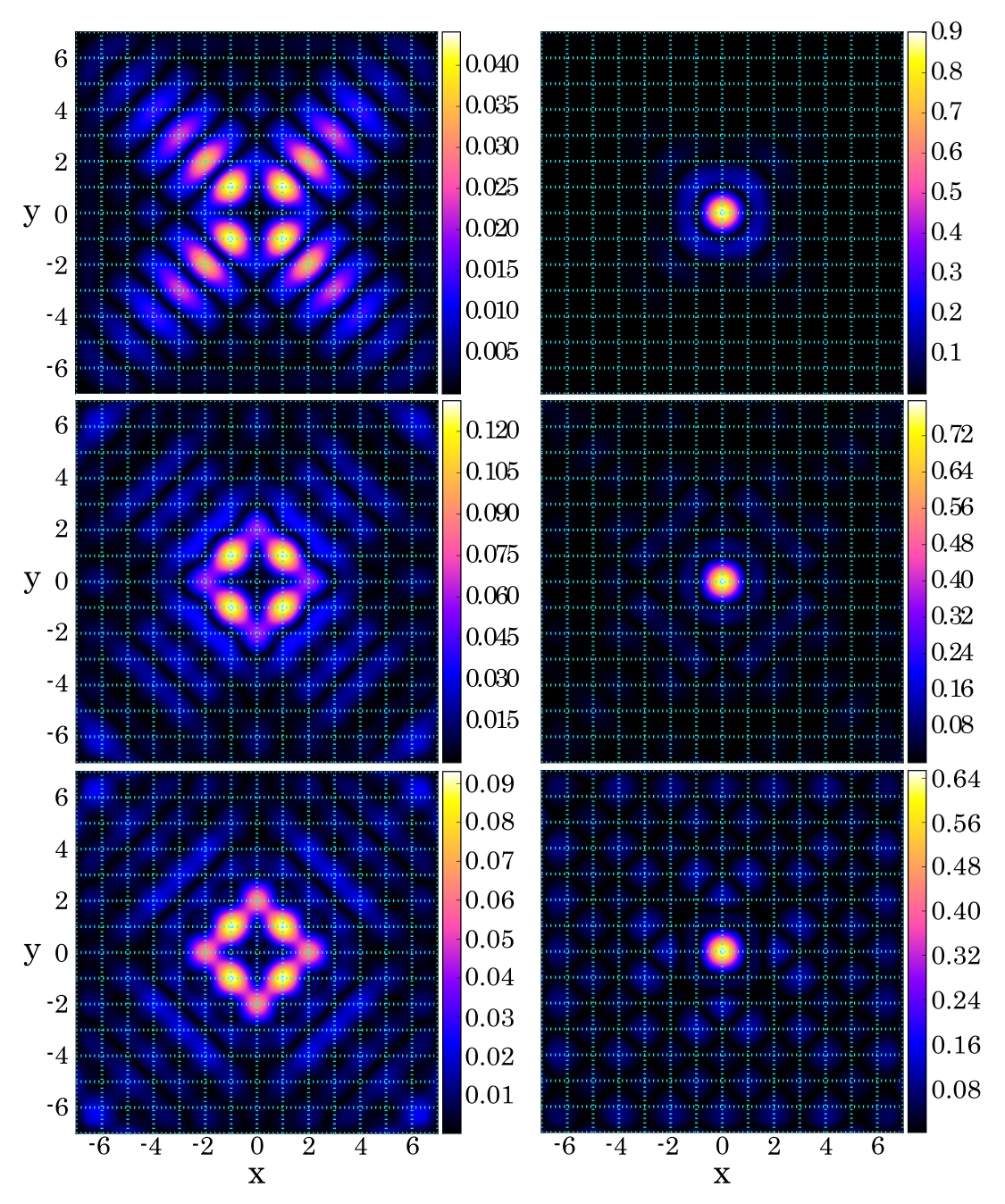}
\caption{Triplet (left) and singlet (right) pair wavefunctions in real-space. From top to bottom $\lambda=0.25, 1.0, 1.5$. The interaction strength is $U/t=-4$ and the system is a $14\times 14$ periodic supercell.
}
\label{fig:psi_r_vs_lbda}
\end{figure}

Figure~\ref{fig:psi_r_vs_lbda} plots, as a function of SOC strength, the behavior
of the pair wavefunctions in real-space, which are given by the Fourier transforms
of $\psi^s_p({\mathbf k})$ and $ \psi^t_p({\mathbf k})$.
For small values of SOC the singlet pair wavefunction (right column) closely resembles the pair wavefunction in the absence of SOC,
with a large well-localized peak indicating strong on-site pairs. However, at increasing values of SOC strength the amplitude
of the central peak is reduced and the pair wavefunction becomes less strongly localized. At large $\lambda$ this delocalization is
evident from the development of additional peaks in the singlet pair wavefunction away from the origin.

The triplet pair wavefunction at small values of SOC shows several peaks along both diagonals, with
the strongest amplitudes located at the second-nearest neighbor sites to the origin. As the SOC strength increases,
the amplitude of the pair wavefunction grows at the third-nearest neighbor sites, finally displaying an intriguing
diamond pattern of peaks at the second- and third-nearest neighbor sites for large SOC strengths. This structure is a 
consequence of the spin-flipping process introduced by the presence of SOC, which implies that for a given spin
at the origin, a parallel spin can be found two nearest-neighbor sites away as a result of successive hops accompanied 
by spin flips.  

\begin{figure}[!ht]
\includegraphics[width=\columnwidth]{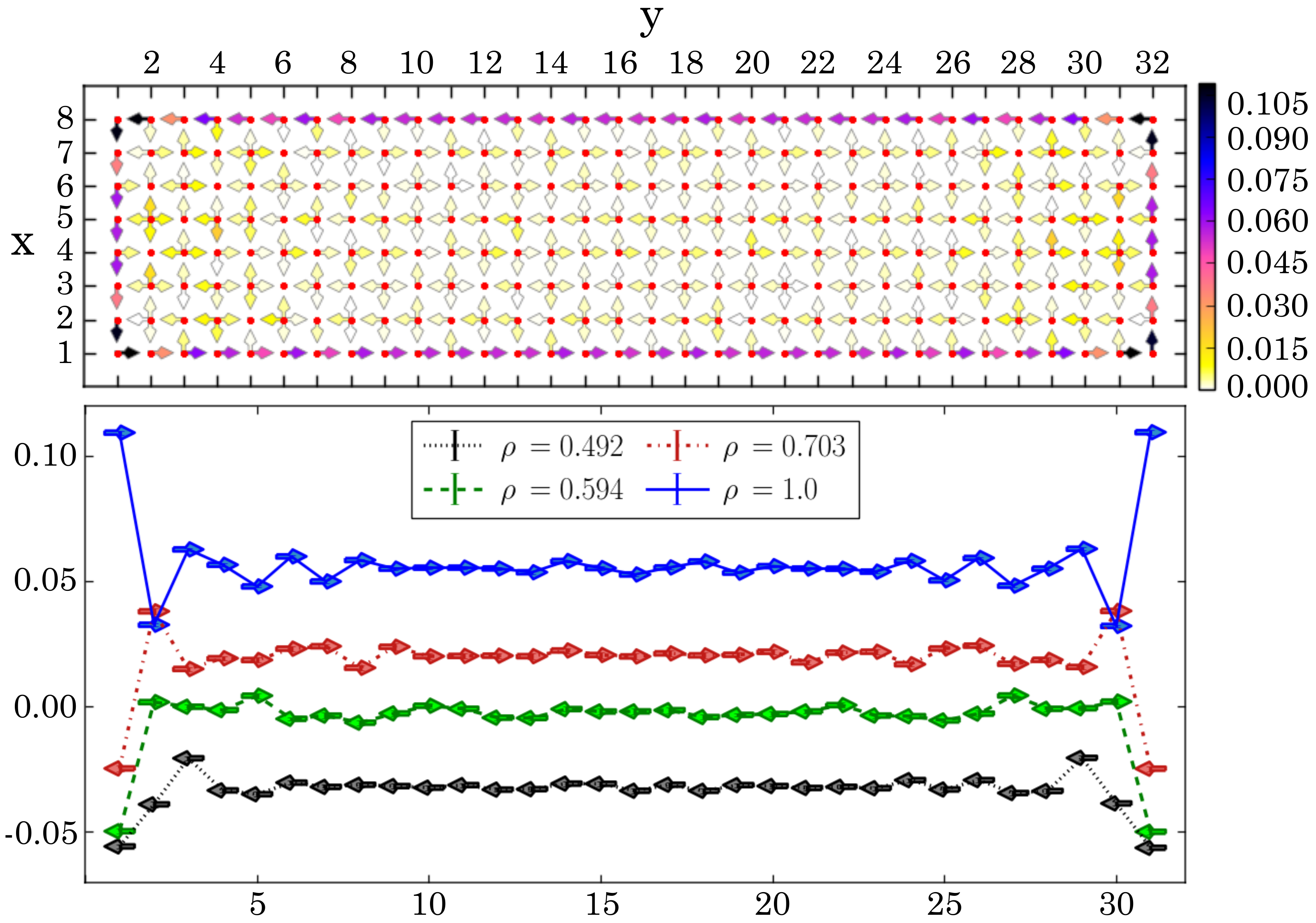}
\caption{Edge current in the open system. The top panel shows the edge current at half-filling for spin-$\uparrow$, with the direction
and magnitude of the current indicated by the arrow direction and its color. The bottom panel 
illustrates the evolution of the current with doping, plotting 
the current in the $y$-direction vs.~$y$ at $x=1$.}
\label{fig:edgec_w_vs_x}
\end{figure}

The potential realization of non-trivial topological and pairing states, with exotic spin transport properties, 
makes this system particularly compelling.  Given that the presence of persistent, topologically protected edge 
modes is an important first step towards reliable spintronic devices and quantum computation, a high-accuracy 
numerical treatment of these behaviors is essential. We examine the existence and behavior of 
edge currents in the system with open boundary conditions. We have verified by comparing multiple calculations 
with periodic and open boundary conditions that the other properties discussed so far remain consistent. In a system with 
open boundaries we compute the current operator, 
\begin{equation}
\mathbf{j}_{nm,\sigma} = -it\left(c^\dagger_{m\sigma}c_{n\sigma} - c^\dagger_{n\sigma}c_{m\sigma}\right),
\end{equation}
which provides a measure of the current for spin-$\sigma$ particles between sites $n$ and $m$.
The top row of Fig.~\ref{fig:edgec_w_vs_x} plots $\mathbf{j}_{nm,\uparrow}$ for $\lambda=2.0$ and clearly establishes the presence
of a current along the boundaries of the system. The strength of this edge current increases monotonically with increasing SOC. 
The current for spin-$\downarrow$ is of equal magnitude and opposite direction, as required by time-reversal symmetry,
such that the net current vanishes in this system.

While the present study is focused on the spin-balanced system at half-filling, it has been suggested 
\cite{2DRSOC_OPLATT_w_Zeeman_MF, Lewenstein2DRSOC_w_Zeeman, PhysRevLett.104.040502} 
that, with the introduction of a time-reversal-symmetry breaking Zeeman field and appropriate parameters, this system 
may host Majorana fermions. 
Mean-field calculations at half-filling indicate that the presence of a 
Zeeman field, even of considerable magnitude, is not sufficient to create spin-$\uparrow$ and spin-$\downarrow$ currents 
of different magnitude, and the system remains in a state with Chern number $0,\pm 2$ 
\cite{Lewenstein2DRSOC_w_Zeeman}.
For doped systems, mean-field studies suggest that the spin-$\uparrow$ and spin-$\downarrow$ currents can have
different magnitudes, and the system can enter topologically protected phases with Chern number $0, \pm 1$. 
Consequently, the most promising parameter regimes to observe non-trivial topological behaviors lie away from half-filling.

We investigate the effect of doping on the edge currents  in the bottom panel of Fig.~\ref{fig:edgec_w_vs_x}.
At half-filling the current has a consistent direction along the edge, with a peak at the corners
(where two edges meet). At intermediate doping the current changes direction along the edge, and a small closed loop
of current forms at each corner of the system. With further doping, the magnitude of the current along the edge
and away from the corners is significantly reduced, however the small loops of current at the corners of the system remain. 
Near quarter-filling the current regains consistent direction, opposite to that at half-filling, and without 
the current loops observed at intermediate doping. 
This behavior is closely connected to the net helicity of the system, which is
determined by the occupation of the helicity bands, $\varepsilon^\pm_\mathbf{k}$. Away from
half-filling, occupation of $\varepsilon^+_\mathbf{k}$ is reduced, resulting in a reduction of the net helicity and
consequently a reduction of the edge current. At larger doping, occupation is limited to $\varepsilon^-_\mathbf{k}$ 
and the net helicity of the system changes sign, indicated by the change of direction of the current at large doping.
The structures in the current are boundary effects of the closed systems.
In semi-periodic systems the current  is essentially constant along the edge.
However a consistent trend in the dependence of the magnitude on doping is observed.

These observations leave open the intriguing question of how the supersolid state we have characterized evolves as a function of 
spin-imbalance and doping. For spin-imbalanced lattice fermions without SOC, the system supports finite-momentum pairing states, 
which is seen directly in Hartree-Fock-Bogoliubov type calculations \cite{FFLO2D}
and can be inferred, via particle-hole symmetry, from more rigorous many-body results on
the doped repulsive Hubbard model, which is predicted by AFQMC
and other calculations \cite{Chia-Chen-sdw,stripe-Hubbard} 
 to support spin-density waves.  Exactly how these FFLO-type pairing states compete or 
coexist with possible Majorana modes in the presence of SOC is a question deserving 
of careful further numerical investigation. Because of broken time-reversal symmetry, the sign problem will reemerge, but can be 
systematically controlled using constrained-path AFQMC \cite{CPMC}. 

In summary we have presented the first numerically exact precision many-body study of a system
directly realizable by ultracold atom experiments in an optical lattice with artificial gauge field. 
The system exhibits exotic properties, with a supersolid phase containing both 
singlet and triplet pairing, and topological signatures in the form of edge currents from SOC. 
Using state-of-the-art quantum Monte Carlo simulations, we have provided a full, unbiased  
treatment of the many-body Schrodinger equation to reliably characterize the effects of 
strong interactions and its interplay with band structure and SOC in the ground state.
We calculated edge currents, which provide a fingerprint for 
 the presence of topological behavior, and are a 
possible precursor of Majorana edge modes when the parameter space of this system is further tuned and 
scanned. 

This research was supported by NSF
(grant no.~DMR-1409510), and the Simons Foundation. 
Computing was carried out at the Extreme Science and Engineering Discovery Environment (XSEDE), 
which is supported by NSF grant number ACI-1053575, and
the computational facilities at the College of William and Mary.

\bibliography{AFQMC_OPLATT}

\end{document}